\documentclass[twocolumn,tighten]{aastex6}

\usepackage{placeins}
\usepackage{natbib}
\usepackage{amsmath}
\usepackage{color}

\begin{document}

\newcommand\be{\begin{equation}}
\newcommand\en{\end{equation}}

\shorttitle{Concentric Rings and Gaps in Protoplanetary Disks} 
\shortauthors{Bae et al.}

\title{ON THE FORMATION OF MULTIPLE CONCENTRIC RINGS AND GAPS IN PROTOPLANETARY DISKS}

\author{Jaehan Bae\altaffilmark{1}, Zhaohuan Zhu\altaffilmark{2}, and Lee Hartmann\altaffilmark{1}}

\altaffiltext{1}{Department of Astronomy, University of Michigan, 1085 S. University Ave., Ann Arbor, MI 48109, USA} 
\altaffiltext{2}{Department of Physics and Astronomy, University of Nevada, Las Vegas, 4505 South Maryland Parkway, Las Vegas, NV 89154, USA}

\email{jaehbae@umich.edu, zhzhu@physics.unlv.edu, lhartm@umich.edu}

\begin{abstract}

As spiral waves driven by a planet in a gaseous disk steepen into a shock, they deposit angular momentum, opening a gap in the disk.
This has been well studied using both linear theory and numerical simulations, but so far, only for the primary spiral arm -- the one directly attached to the planet.
Using two-dimensional hydrodynamic simulations, we show that the secondary and tertiary arms driven by a planet can also open gaps as they steepen into shocks.
The depths of the secondary/tertiary gaps in surface density grow with time in a low viscosity disk ($\alpha = 5\times10^{-5}$), so even low-mass planets (e.g., super-Earth or mini-Neptune-mass) embedded in the disk can open multiple observable gaps, provided that sufficient time has passed. 
Applying our results to the HL Tau disk, we show that a single 30 Earth-mass planet embedded in the ring at 68.8~au (B5) can reasonably well reproduce the positions of the two major gaps at 13.2 and 32.3 au (D1 and D2), and roughly reproduce two other major gaps at 64.2 and 74.7 au (D5 and D6) seen in the mm continuum.
The positions of secondary/tertiary gaps are found to be sensitive to the planetary mass and the disk temperature profile, so with accurate observational measurements of the temperature structure the positions of multiple gaps can be used to constrain the mass of the planet.
We also comment on the gaps seen in the TW Hya and HD~163296 disk.

\end{abstract}


\section{INTRODUCTION}

Recent high-resolution observations at (sub)-mm and $\mu$m wavelengths have revealed interesting structures in protoplanetary disks, including rings and gaps (e.g., HL~Tau, \citealt{alma15}; TW~Hya, \citealt{andrews16}, \citealt{tsukagoshi16}, \citealt{vanboekel17}; HD~163296, \citealt{isella16}, \citealt{zhang16}; HD~97048, \citealt{ginski16}, \citealt{walsh16}), spiral waves (e.g., MWC~758, \citealt{grady13}, \citealt{benisty15}; SAO~206462, \citealt{muto12}, \citealt{garufi13}, \citealt{stolker16}, AB Aug, \citealt{tang17}), and vortex-like asymmetries (e.g., Oph~IRS~48, \citealt{vandermarel13}).
Among these structures, particularly surprising is rings and gaps, not simply because of their existence but because of their concentric and axisymmetric shape, seeming prevalence among the young protoplanetary disk population (although based on a small number of objects observed with sufficient spatial resolution), and multiplicity in individual disks.

The origin of the observed concentric rings and gaps is yet to be confirmed.
Possible mechanisms include gravitational interaction between planets and disks \citep[e.g.,][]{krist00,calvet02,vandermarel15,dong17}, rapid changes in dust properties at the condensation front of volatile species \citep{zhang15,debes16}, variation of magnetic activity within disks \citep{johansen09,ruge13,flock15,pinilla16,ruge16} or in MHD disk wind \citep{suriano17}, dust evolution \citep{birnstiel15,okuzumi16}, photoevaporation \citep{gorti11,owen12}, secular gravitational instability \citep{takahashi14,takahashi16}, and self-induced dust traps \citep{gonzalez15,gonzalez17}.

Assuming planets are responsible for the disk gaps, numerical simulations generally use as many planets as the number of gaps to explain multi-ring/gap structures (e.g., \citealt{dipierro15}, \citealt{isella16}, \citealt{jin16}).
In some simulations, however, it appears that multiple gaps and rings can form from one planet (see for example Figure 1 and 4 of \citealt{zhu14}, and Figure 5 of \citealt{bae16b}).
Interestingly, the additional gaps open far from the planet, $\sim0.6~r_p$ in \citet{zhu14} and $\sim0.45$ and $0.6~r_p$ in \citet{bae16b}, where $r_p$ denotes the planet's semi-major axis.
\citet{dong17} recently showed that an observable ``double gap'' can be opened at the vicinity of a super-Earth mass planet, about a disk scale height from the planet, in a low-viscosity disk.
In addition to the ``double gap'' adjacent to the planet, in \citet{dong17} an additional gap opens at $\sim 0.6~r_p$ which seems to share the same origin with the ones seen in \citet{zhu14} and \citet{bae16b}.  
The detailed cause of the additional sets of rings and gaps that form far from a planet, however, has not yet been investigated in the literature.

In this paper, we study the origin of multiple rings and gaps created by a single planet.
When a planet forms, the gravitational interaction between the planet and its host disk excites spiral arms.
Even for a sufficiently low-mass planet, for which the excitation and initial propagation of spiral density waves launched by the planet can be explained well by linear theory, it is known that spiral waves non-linearly steepen to spiral shocks as the waves propagate \citep{goodman01}.
When a spiral wave shocks, angular momentum is transferred to the disk, which in turn leads the redistribution of disk material, opening a gap \citep{rafikov02}.
This phenomenon has been well studied using both linear theory \citep{goodman01,rafikov02} and numerical simulations \citep{muto10,dong11,duffell12,zhu13,dong17}, but so far, only for the primary arm -- the one directly attached to the planet.

Previous protoplanetary disk simulations have shown that a secondary spiral arm can be excited by a planet \citep[e.g.,][]{kley99,fung15,juhasz15,zhu15,bae16aa,bae16b,lee16}.
In addition, some simulations have revealed that a tertiary arm can also be excited in the inner disk \citep{zhu15,bae16b}.
We will present the mechanism for multiple spiral arms in a forthcoming paper (Bae \& Zhu 2017, in preparation).
In the present work, we show that secondary and tertiary spiral arms excited by a planet can create secondary and tertiary gaps in the inner disk when they shock, in the same way that the primary arm opens a gap.
We propose that spiral shocks associated with secondary/tertiary arms can be a possible origin for the observed multiple sets of rings and gaps in some protoplanetary disks, as we will show for the HL Tau disk as an example later in the paper.

This paper is organized as follows.
In Section \ref{sec:method} we introduce our numerical setup.
In Section \ref{sec:results} we present our results, describing the formation of multiple rings and gaps from a low-mass planet in Section \ref{sec:01mth} and a high-mass planet in Section \ref{sec:3mth}.
We then study a long-term evolution of the rings and gaps in Section \ref{sec:long_term}.
In Section \ref{sec:hltau}, applying our results we explain the multiple rings and gaps seen in the HL Tau disk using a single planet.
Finally, in Section \ref{sec:discussion} we discuss implications of the present work and comment on some individual disks with multiple rings and gaps.

\section{NUMERICAL METHODS}
\label{sec:method}

We solve the hydrodynamic equations for mass and momentum conservation in two-dimensional polar coordinates $(r,\phi)$ using FARGO 3D \citep{benitez16}:
\be\label{eqn:mass}
{\partial \Sigma \over \partial t} + \nabla \cdot (\Sigma v) = 0,
\en
\be\label{eqn:momentum}
\Sigma \left( {\partial v \over \partial t} + v \cdot \nabla v \right) = - \nabla P - \Sigma \nabla (\Phi_* + \Phi_p).
\en
In the above equations, $\Sigma$ is the gas surface density, $v$ is the velocity, $P=\Sigma c_s^2$ is the pressure where $c_s$ is the isothermal sound speed, and $\Phi_* = GM_*/r$ is the gravitational potential from the central star.
The planetary potential $\Phi_p$ is
\be
\label{eqn:potential}
\Phi_p (r, \phi)= -{{GM_p} \over {(|{\bf{r}}-{\bf{r_p}}|^2 + b^2)^{1/2}}},
\en
where $M_p$ is the planetary mass, ${\bf r}$ and ${\bf r_p}$ are the radius vectors of the center of grid cells in question and of the planet, and $b$ is the smoothing length which is assumed to be $10\%$ of the disk scale height at $r=r_p$. 
To provide a clearer picture, we ignore the indirect term that arises from the offset between the central object and the origin of the coordinate system; however, we confirmed that including the indirect term does not affect our conclusion. 

We increase the planetary mass over a period time $t_{\rm growth}$ at the beginning of each simulation in order to avoid sudden perturbation to the disk, following $M_p(t) \propto \sin[(\pi/2) {\rm min}(t/t_{\rm growth}, 1)]$.
The growth timescale $t_{\rm growth}$ is set to $t_{\rm growth} = 20~t_p \times (M_p/ 0.1M_{\rm th})$ for various planetary masses, with a maximum of $100~t_p$, where $t_p=2\pi/\Omega_p$ is the planet's orbital time.
In the above growth timescale equation, $M_{\rm th} \equiv c_s^3/\Omega G = M_* (h/r)_p^3$ is the so-called thermal mass \citep{lin93,goodman01}, at which planetary mass the Hill radius of the planet is comparable to the disk scale height.
Below the thermal mass, the excitation and the initial propagation of the spiral density wave from a planet can be well approximated in the linear regime \citep[e.g.,][]{goldreich80}.
When planetary mass becomes greater than the thermal mass, however, the spiral density wave becomes non-linear and shock background disk gas as soon as they launch \citep{goodman01}.

Our initial disk has power-law surface density and temperature distributions:  
\be
\label{eqn:init_den}
\Sigma(r)   =  \Sigma_p \left({r \over r_p}\right)^{-p}
\en
and 
\be
\label{eqn:temp}
T(r) = T_p \left( {r \over r_p} \right)^{-q},
\en 
where $\Sigma_p$ and $T_p$ are the surface density and temperature at the location of the planet $r=r_p$.
The above temperature profile corresponds to the disk aspect ratio profile of 
\be
{h \over r} = \left( {h \over r} \right)_p \left( {r \over r_p} \right)^{(-q+1)/2}.
\en 
We adopt $p=1$, $q=0$, and $(h/r)_p = 0.07$ in our fiducial model.

While the assumed constant temperature ($q=0$) may not be a very realistic temperature structure for global protoplanetary disks, this enables us to directly compare our numerical results with predictions for spiral wave excitation and propagation from linear theory that is developed assuming a constant temperature \citep[e.g.,][]{goodman01}.
However, we note that the physics behind the gap opening through spiral shocks should remain the same even for different power-law temperature slopes.
We add a small viscosity $\alpha=5\times10^{-5}$ to the gas, where $\alpha$ denotes a Shakura-Sunyaev viscosity parameter \citep{shakura73}: $\nu = \alpha c_s^2/\Omega$.
As viscosity tends to smooth out structure, we initially choose this low value of $\alpha$ to show the mechanism of gap opening more clearly and to compare the numerical results more directly with linear theory.  
The effects of larger viscosity are discussed in Section \ref{sec:discussion}.

Our simulation domain extends from $r_{\rm in} = 0.2~r_p$ to $r_{\rm out} = 2.0~r_p$ in radius and from 0 to $2\pi$ in azimuth.
Because we use an isothermal equation of state and do not include disk self-gravity, no physical units are necessary and the quantities presented in this work can be scaled.
For our fiducial calculations presented in Section \ref{sec:results}, we adopt 4096 logarithmically-spaced grid cells in the radial direction and 11160 uniformly-spaced grid cells in the azimuthal direction.
For the long-term calculations in Section \ref{sec:long_term} we use $2048 \times 5580$ grid cells.
With both choices all grid cells are nearly square-shaped ($\Delta r : r \Delta\phi \simeq 1:1$).
We have tested with various numerical resolution and found that this very high spatial resolution is required to properly capture the opening of secondary and tertiary gaps.
For the model disk assumed in this work, using $2048 \times 5580$ or more grid cells can resolve the spiral shock at all radii but using $1024 \times 2290$ grid cells fails to resolve the shock in the inner disk $r \lesssim 0.4~r_p$.
When a spiral shock is under-resolved we find that no gap associated with the spiral arm forms.

At the radial boundaries, we adopt a wave-damping zone \citep{devalborro06} to reduce wave reflection.
We have tested with other boundary conditions, including outflow boundary condition and reflecting boundary condition, and confirmed that the choice of boundary condition does not affect our conclusion that secondary/tertiary gaps open when secondary/tertiary arms steepen into shocks.

%figure 1
\begin{figure*}
\centering
\epsscale{1.18}
\plotone{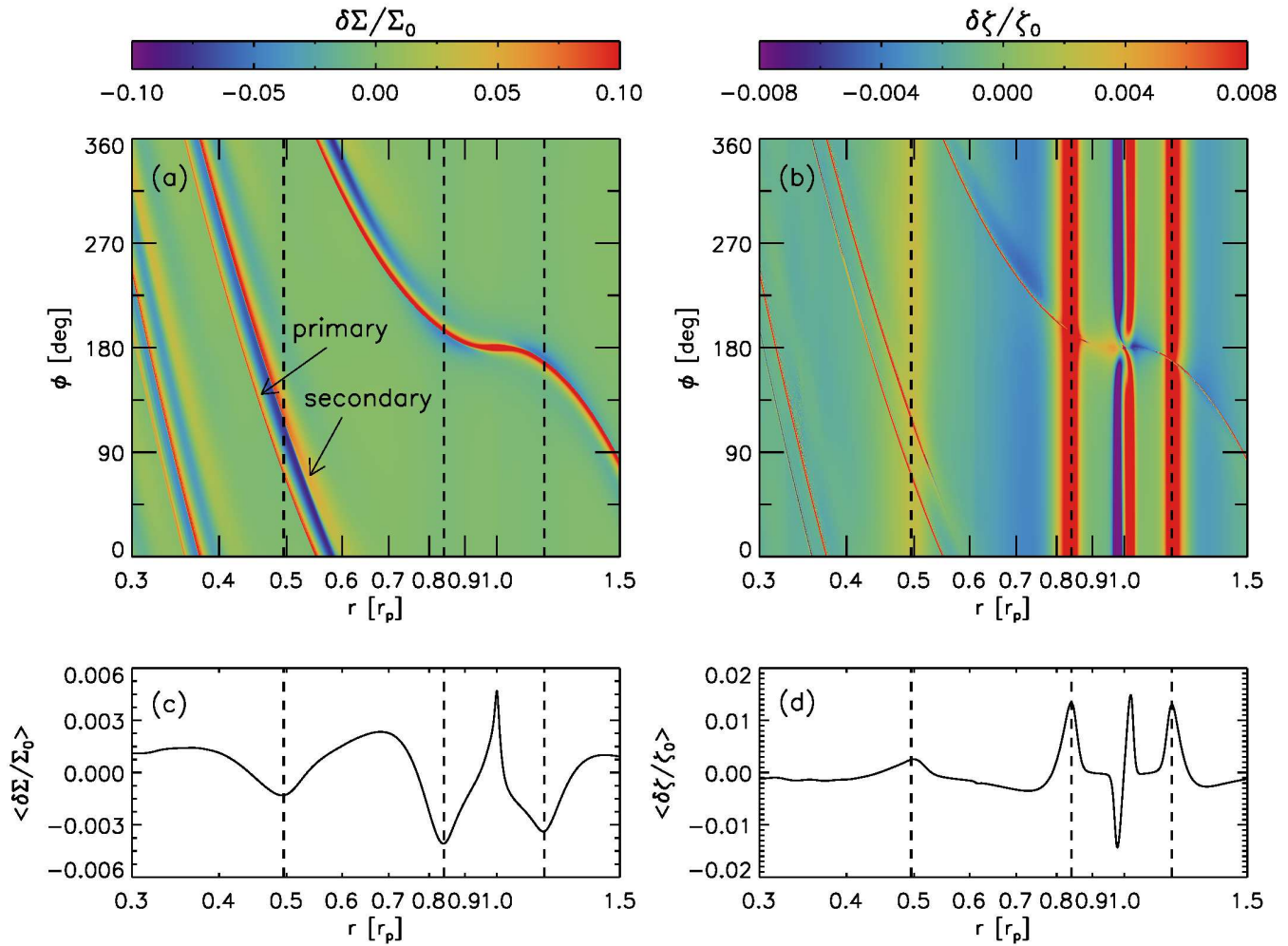}
\caption{The two-dimensional distributions of (a) the perturbed surface density $\delta\Sigma/\Sigma_0$ and (b) the perturbed PV $\delta\zeta/\zeta_0$ at $t=50~t_p$ with a $0.1~M_{\rm th}$ planet. The planet is at $(r,\phi)=(1r_p, 180^\circ)$. The bottom panels present the azimuthally averaged profiles of (c) $\delta\Sigma/\Sigma_0$ and (d) $\delta\zeta/\zeta_0$. In all panels, the vertical dashed lines indicate the local minima in the density distribution at $r = 0.495, 0.84$, and $1.17~r_p$. Note that the locations of these density deficits match very well with the PV local maxima in panels (b) and (d).}
\label{fig:pv_01mth}
\end{figure*}

\section{Formation of Rings and Gaps}
\label{sec:results}

\subsection{Results with a $0.1~M_{\rm th}$ planet }
\label{sec:01mth}

We start with results for a $0.1~M_{\rm th}$ planet (corresponding to $11.4~M_\earth$ with $M_*=1M_\odot$ and $(h/r)_p=0.07$), for which the excitation and the initial propagation of the spiral density wave can be explained with linear theory. 
In Figure \ref{fig:pv_01mth} (a), we show the perturbed density distribution $\delta\Sigma/\Sigma_0$ at $t=50~t_p$, where $\Sigma_0$ is the initial density at each position of the disk and $\delta\Sigma = \Sigma - \Sigma_0$.
The planet launches a trailing spiral arm that is directly attached to the planet (i.e., primary arm). 
The planet starts to launch a secondary arm around $r=0.6~r_p$.
The secondary arm is closely located to the primary in azimuth, only $\sim 50^\circ$ apart.
There is a hint of another arm at $r \sim 0.3 - 0.4~r_p$.
However, this tertiary arm-like structure does not produce a shock within our simulation domain.
We have tested this with a factor of two higher resolution ($8192 \times 22320$ grid cells) but no shock was found to be produced from this arm-like structure.
We find that the perturbation from this tertiary arm-like structure gradually increases but also excites closer to the planet when we increase the planetary mass, as we discuss in the following section.

%figure 2
\begin{figure*}
\centering
\epsscale{1.18}
\plotone{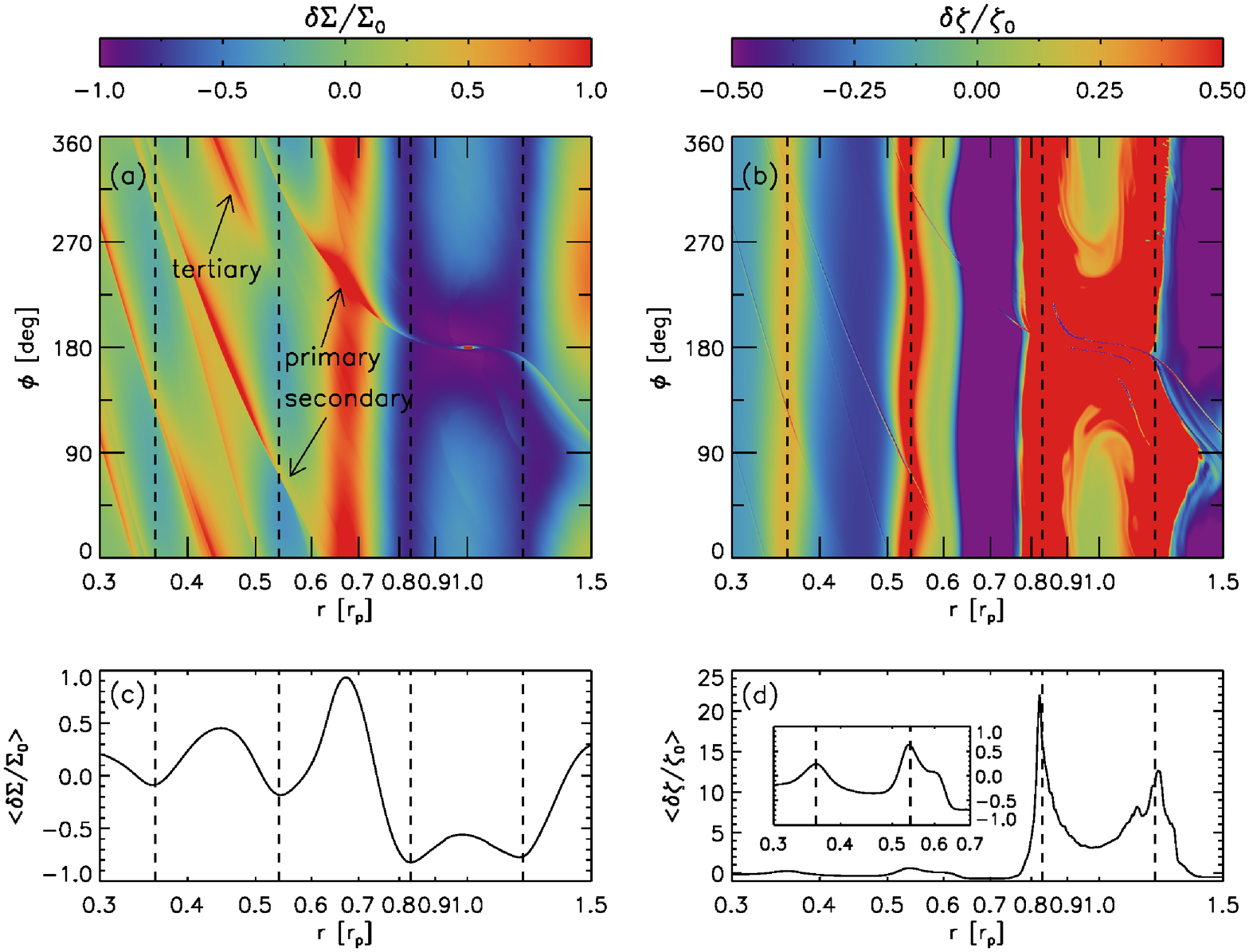}
\caption{Same as Figure 1, but results with a $3~M_{\rm th}$ planet $t=230~t_p$. In order to emphasize the perturbed PV distribution around the gaps associated with the secondary and tertiary arms, $0.75 r_p \lesssim r \lesssim 1.3 r_p$ is saturated purposely in panels (b). In panel (d), the inset presents an expanded view for $0.3 \leq r \leq 0.7$ to more clearly show the perturbed PV by the secondary and tertiary arms. When calculating the azimuthally averaged density perturbation in panel (c), planet's Hill sphere region is excluded. The vertical dashed lines indicate the local minima in the density distribution at $r = 0.36, 0.54, 0.83$, and $1.20~r_p$. Again, note that the locations of the PV local maxima match well with the density gaps.}
\label{fig:pv_3mth}
\end{figure*}

In order to diagnose the radial location at which a planet-driven spiral wave becomes a shock, we compute the potential vorticity (PV) $\zeta$, defined as 
\be
\zeta \equiv {{\nabla \times {\bf v}} \over {\Sigma}}.
\en
For a barotropic fluid as we have in our models, it is well known that the PV is conserved in the absence of shocks.
When a shock is present, on the other hand, the PV experiences a jump at the shock front \citep{li05,dong11}.
In Figure \ref{fig:pv_01mth} (b), we show the perturbed PV distribution $\delta\zeta/\zeta_0$, where $\delta\zeta = \zeta - \zeta_0$ and $\zeta_0 = (\nabla \times {\bf v}_0)/\Sigma_0$ is the unperturbed PV at the beginning of the simulation.
Looking at the perturbed PV distribution along the primary arm first, one can see that it does not produce a shock at the immediate vicinity of the planet but a few scale heights away from the planet as expected from linear theory.
Linear waves from a low-mass planet ($M_p \ll M_{\rm th}$) steepen into shocks as they propagate away from the planet \citep{goodman01} and the distance at which linear waves become a shock can be approximated from linear theory \citep{goodman01}:
\be
| l_{\rm sh} | \approx 0.93 \left( {{\gamma+1} \over 12/5} {M_p \over M_{\rm th}} \right)^{-2/5} h,
\en
where $\gamma$ is the adiabatic index and $h$ is the disk scale height.
With $M_p=0.1~M_{\rm th}$ and $(h/r)_p=0.07$, $l_{\rm sh} \approx \pm 2.5h \approx \pm 0.176$, so our numerical results show a good agreement with the prediction from linear theory.
The PV perturbation accumulates over time, creating two rings -- one interior and the other exterior to the planetary orbit -- of high PV region around the radii a spiral becomes a shock ($r \sim 0.84$ and $1.17~r_p$).

We find that the same happens for the secondary arm.
The secondary arm starts to shock background medium around $r=0.5~r_p$, producing a PV jump at its shock front and creating a ring of high PV.

In Figure \ref{fig:pv_01mth} (c) and (d), we plot the azimuthally averaged profiles of the perturbed density and perturbed PV.
As can be seen from the figures, the locations where spiral arms produce shocks and the locations of density deficits match very well with each other.
Again, no apparent shock is seen at the tertiary arm-like structure so no gap forms from it.

In the outer disk, additional spiral arms are not excited within our simulation domain ($r_{\rm out}=2~r_p$), so only one gap associated with the primary arm opens at $r=1.17~r_p$.
As an aside, for the planetary mass and disk temperature profile considered here ($0.1~M_{\rm th}$ and $(h/r)_p=0.07$), it is unlikely that additional spiral arms form even beyond $r=2~r_p$ (Bae \& Zhu 2017, in preparation).

\subsection{Results with a $3~M_{\rm th}$ planet }
\label{sec:3mth}

We now describe results for a more massive planet, $3~M_{\rm th}$ (corresponding to $1.03~M_{\rm Jup}$ with $M_*=1M_\odot$ and $(h/r)_p=0.07$).
This mass is chosen such that three spiral arms and the associated gaps develop well within our simulation domain.
As we gradually increase the planetary mass from $0.1~M_{\rm th}$ to $3~M_{\rm th}$, we find that the secondary and tertiary arms are excited in the inner disk closer to the planet -- this is much more prominent for the tertiary as we show below.
Increasing planetary mass also increases the surface density perturbations at the shock front and the separation between the arms.

In Figure \ref{fig:pv_3mth} (a), we plot the perturbed density distribution.
Three arms are launched by a $3~M_{\rm th}$ planet: a primary arm that is attached to the planet, a secondary arm launched around $r=0.6~r_p$, and a tertiary arm launched around $r=0.5~r_p$.
While the secondary arm launches at a similar radius to the low-mass planet case, the tertiary arm starts to appear closer to the planet, presumably because non-linear effects shift the launching radius of the tertiary arm more than that of the secondary arm (Bae \& Zhu 2017, in preparation).
Also, the arm-to-arm separation increases as the planetary mass increases, consistent with previous studies (e.g., \citealt{zhu15,fung15,lee16}).
At $r=0.6~r_p$, the secondary arm is offset by $112^\circ$ from the primary arm, as opposed to $50^\circ$ in the low-mass planet case. 
Another noticeable difference is that the planet opens a deep gap around itself and therefore it is no more embedded in the disk.

In Figure \ref{fig:pv_3mth} (b) we plot the perturbed PV distribution to find the shock locations.
The secondary and tertiary arms start to  shock at $r \sim 0.56$ and $0.37~r_p$.
The arms accumulate PV as they shock, creating rings of high PV. 
The azimuthally averaged distributions of the perturbed density and PV are presented in Figure \ref{fig:pv_3mth} (c) and (d), respectively.
As in the case of the low-mass planet, the locations of density gaps and the local PV maxima match well with each other, supporting that the idea of spiral shocks opening gaps is also valid for more massive gap-opening planets.

%figure 3
\begin{figure*}
\centering
\epsscale{1.17}
\plotone{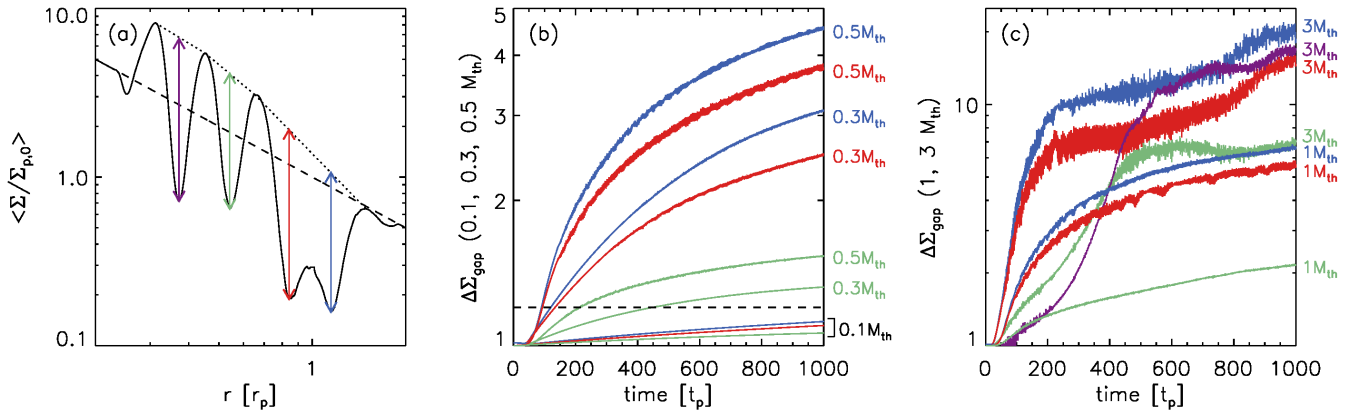}
\caption{(a) The azimuthally averaged surface density distribution for a $3~M_{\rm th}$ case at $t=500~t_p$, normalized by the initial surface density at the planet's position $\Sigma_{p,0}$, to provide a view of our gap depth calculation. The black dotted lines are power-law fits for the rings adjacent to the gaps and the arrow indicates the gap depth $\Delta\Sigma_{\rm gap}$, which we define as the ratio between the expected density assuming the power-law fit and the actual density. The black dashed line indicates the initial density distribution. (b) The gap depth $\Delta\Sigma_{\rm gap}$ for 0.1, 0.3, and $0.5~M_{\rm th}$ planets. Different colors present: (blue) primary gap in the outer disk, (red) primary gap in the inner disk, (green) secondary gap, and (purple) tertiary gap as illustrated in panel (a). The horizontal dashed line indicates $\Delta\Sigma_{\rm gap} = 1.2$ (see Section \ref{sec:discussion}). (c) Same as panel (b) but for 1 and $3~M_{\rm th}$ planets. In panel (a), the density deficit near the inner boundary at $r \sim 0.25~r_p$ is not due to spiral shocks but is a feature produced by the implementation of the damping zone. In the damping zone physical quantities (i.e., density, velocity) relax toward their initial values over some timescale, so this effectively applies some viscosity and creates the gap-like feature.}
\label{fig:long_term}
\end{figure*}

\section{Long-term Evolution of Rings and Gaps}
\label{sec:long_term}

One of the interesting questions regarding the secondary/tertiary gaps is how deep could they be.  
To address the question, we ran simulations for 1000 planetary orbits and measured the depths of gaps for various planetary masses: 0.1, 0.3, 0.5, 1, 2, and $3~M_{\rm th}$, corresponding to $11.4~M_\earth, 0.10~M_{\rm Jup}, 0.17~M_{\rm Jup}, 0.34~M_{\rm Jup}, 0.69~M_{\rm Jup}$, and $1.03~M_{\rm Jup}$ when $M_*=1M_\odot$ and $(h/r)_p=0.07$ are assumed.

We computed the gap depth $\Delta\Sigma_{\rm gap}$ as follows.
We first generated a power-law fit for the adjacent density rings to a gap, illustrated with dotted lines in Figure \ref{fig:long_term} (a).
For the inner and outer primary gaps (indicated with red and blue arrows in the figure), we generated the fit using the density rings at the inner and outer gap edges, instead of the local density maximum at the corotation region ($r \sim 1~r_p$).  
We then calculated the expected surface density at the gap center (i.e., local density minimum) using the power-law fit.
The gap depth is defined as the ratio between the power-law surface density at the gap center from the fit and the actual surface density there.
This ratio should be unity in a power-law disk when there is no gap, and gradually increase as a gap opens.

The gap depth calculated following this method is plotted in Figure \ref{fig:long_term} (b) and (c).
Looking first at planets with $M_p \leq 1M_{\rm th}$, neither the primary nor the secondary gaps reach a steady state until $1000~t_p$.
We find that the two primary gaps have a comparable depth, although the gap exterior to the planetary orbit is slightly deeper than the gap interior to the planetary orbit in all cases.
The secondary gaps are shallower than the primary gaps by a few percent ($0.1~M_{\rm th}$) to a factor of $\sim3$ ($1~M_{\rm th}$) in terms of $\Delta\Sigma_{\rm gap}$ value.
Secondary gaps are shallower than primary gaps presumably because secondary arms are weaker than primary arms when compared at their shock locations.

For the $3~M_{\rm th}$ case, we find that the gap depths reach to a quasi-steady state as the Rossby wave instability (RWI) is triggered at the rings.
This happens at $\sim200~t_p$ for the primary rings, $\sim500~t_p$ for the secondary ring, and $\sim600~t_p$ for the tertiary ring.
On the other hand, as inferred from Figure \ref{fig:long_term} (a) and (c), the RWI is not able to get rid of the rings/gaps but it only impedes further growth of the rings/gaps.
It is also interesting to note that the tertiary gap can be as deep as or even deeper than the primary/secondary gaps.
We conjecture that this is because (1) the tertiary arm can be as strong as the secondary arm at least locally in radius (Bae \& Zhu 2017, in preparation) and/or (2) the gaps are (partially) filled back via the RWI in such a way that the resulting tertiary gap depth is comparable to or deeper than the primary/secondary gaps. 

\begin{figure*}
\centering
\epsscale{1.1}
\plotone{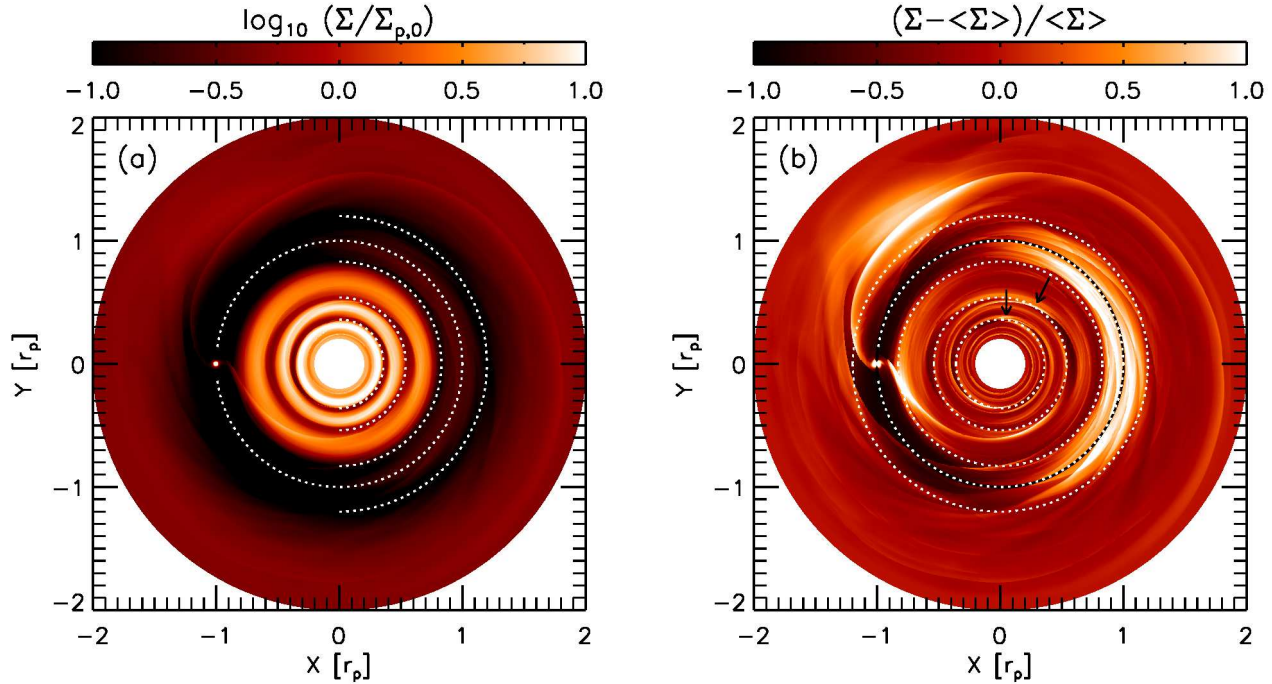}
\caption{(a) The surface density distribution at $t=1000~t_p$ with a $3~M_{\rm th}$ planet, normalized by the initial surface density at the planet's position, in a logarithmic scale. The dotted full circle in white indicates the planetary orbit while the dotted half circles in white show the gaps to highlight their concentric shape: from inner to outer disk, the gaps associated with the tertiary arm, secondary arm, primary arm interior to the planetary orbit, and exterior to the planetary orbit. 
(b) The perturbed density above/below the azimuthal average at each radius is shown to highlight non-axisymmetric structures. The dotted white circles indicate the planetary orbit and the gap locations. The two arrows indicate the locations where the (right arrow) secondary and (left arrow) tertiary arms shock disk gas.}
\label{fig:density}
\end{figure*}

In Figure \ref{fig:density} (a) we show the surface density distribution at $t=1000~t_p$ for the $3~M_{\rm th}$ case.
Multiple sets of rings and gaps are immediately noticeable: one gap exterior to the planetary orbit opened by the primary arm, three gaps interior to the planetary orbit each of which is opened by the primary, secondary, and tertiary arms (from large to small radii).
As depicted in the figure all the gaps and rings are concentric, similar to the ones seen in recent high resolution observations.
In Figure \ref{fig:density} (b) we display the density perturbation above/below the azimuthal average at each radius in order to highlight any non-axisymmetric structure present in the disk.
One can see that the rings are nearly axisymmetric, which is again similar to the observed multiple rings.
Note also that the radial locations where the PV experiences a jump (indicated by the arrows in the figure) match well with the gap locations.

%figure 4

\section{Parameter Study and Possible Application to HL Tau}
\label{sec:hltau}

Now that we have shown that multiple gaps can be opened by multiple spiral waves from a single planet, we explore the dependence of additional gap positions on the background disk temperature profile and the planetary mass, and consider a possible application of our results to the case of HL Tau.
Our goal here is not to produce a detailed model of the disk but to explore whether some of the dust gaps might be explained with a single planet.

The dust emission of the HL Tau disk shows remarkable multiple, concentric rings and gaps \citep{alma15}.
The ALMA observation revealed four major gaps at 13.2 (D1), 32.3 (D2), 64.2 (D5), and 73.7~au (D6), and rings in between \citep{alma15}.
Also, there are additional narrower and shallower rings and gaps (B3/D3, B4/D4, and B7/D7; \citealt{alma15}).

To model the HL Tau disk, we adopt the gas surface density profile constrained by the mm continuum observations of \citet{kwon11}:
\be
\label{eqn:init_den}
\Sigma(r)   =  \Sigma_c \left({r \over r_c}\right)^{-\gamma}\exp\left[-\left( {r \over r_c}\right)^{2-\gamma}\right]
\en
where $\gamma=-0.23$ and $r_c=79$~au.
For the temperature profile, we make use of the brightness temperature obtained in \citet{alma15}: using 59~K at 20~au and 24~K at 81~au we obtain the temperature power-law index $q=0.65$\footnote{Since the HL Tau disk is likely to be (marginally) optically thick at 1.3~mm \citep{alma15,carrasco16,pinte16}, this temperature profile may not be a completely accurate estimate of the midplane temperature.}.
Assuming a central stellar mass of $1.3~M_\odot$ \citep{alma15} and a mean molecular weight of 2.4, the disk aspect ratio profile follows $h/r = 0.074(r/r_p)^{0.175}$, consistent with irradiation heating by the central object in a moderately-flared disk.
For the simulations presented in this section, we adopt $\alpha=3\times10^{-4}$ as estimated for the HL Tau disk by \citet{pinte16}.
We adopt $r_{\rm in} = 0.05~r_p$ and $r_{\rm out} = 2~r_p$ and use $4096\times6974$ grid cells.

Assuming that the planet has a circular orbit and does not migrate, we place a planet at B5 (68.8~au; \citealt{alma15}), the ring in between D5 and D6, which is also suggested by the ``double gap'' feature in \citet{dong17}.
We then vary the planetary mass and search for the one that opens secondary and tertiary gaps at the locations of the observed D1 and D2.
Since we have only the gas component in our simulations while the ALMA continuum emission traces solid particles, we do not attempt to reproduce the depth of each gap in the observed intensity profile.
As we show below, we note that the planetary mass that best reproduces the two innermost gaps (D1 and D2) does not result in a perfect match for the gaps near the planet (D5 and D6).
However, we do not aim to reproduce the locations of D5 and D6, since 
the gas flow near the planetary orbit can be extremely complicated \citep[e.g.,][]{fungetal15,fung16} and thus the dust distribution in the corotation region, which produces the observed emission, cannot be accurately predicted with gas-only calculations.
More thorough comparisons between numerical models and the observed data will require multi-fluid calculations that can simultaneously compute gas and dust dynamics; we defer this to a future study.
On the other hand, we note that the locations of the secondary/tertiary gaps do not evolve over time when the temperature profile and the location of the planet is fixed as we assumed, and the gap depths are likely to grow over time as Figure \ref{fig:long_term} suggests.

%figure 5
\begin{figure*}
\centering
\epsscale{1.15}
\plotone{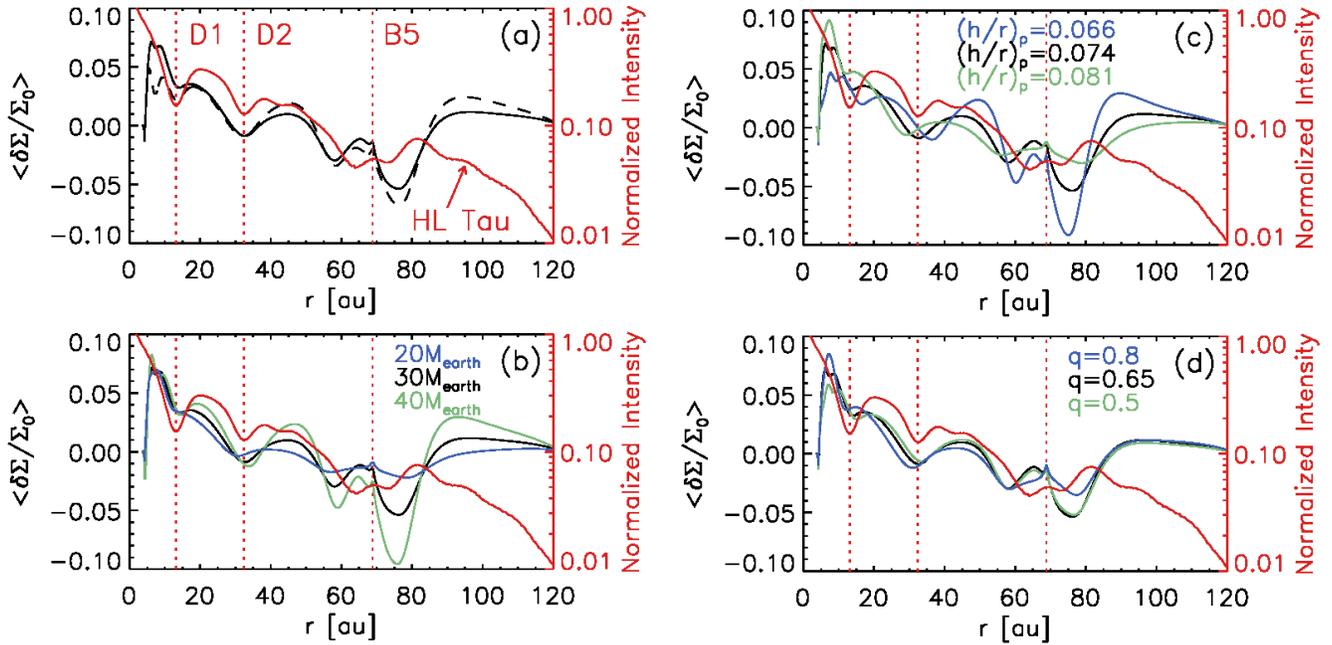}
\caption{(a) The black curve presents the azimuthally averaged perturbed density $\langle \delta\Sigma / \Sigma_0 \rangle$, assuming $M_p=30~M_\earth$, $(h/r)_p =0.074$, and $q=0.65$. The black dashed curve presents $\langle \delta\Sigma / \Sigma_0 \rangle$, divided by two, when $M_p=60~M_\earth$, $(h/r)_p =0.081$, and $q=0.65$ are used. The red curve shows the normalized intensity distribution of the HL Tau disk at Band 6 \citep{alma15}. The vertical red dotted lines indicate the locations of the two innermost gaps seen in the ALMA data (D1 and D2) and the ring at 68.8~au where we place the planet (B5). The azimuthally averaged perturbed density distributions using different (b) planetary masses, (c) disk temperature, and (d) temperature slope. All simulation data shows results at $t=100~t_p$. Unless otherwise stated in each panel, we use our fiducial parameters: $M_p=30~M_\earth$, $(h/r)_p =0.074$, and $q=0.65$.}
\label{fig:hltau_param}
\end{figure*}

Based on a suite of calculations varying the planetary mass we find that the two innermost gaps can be reasonably well reproduced with a 30 Earth-mass planet at B5.
In Figure \ref{fig:hltau_param} (a) we present the azimuthally averaged perturbed density distribution, together with the normalized intensity profile of the ALMA continuum observation at Band 6 \citet{alma15}.
The planet opens three gaps interior to its orbit, each of which is associated with the primary, secondary, and tertiary spiral arm (from large to small radii), and one gap exterior to its orbit.
As shown, with the 30 Earth-mass planet at 68.8~au the locations of the secondary and tertiary gaps from our simulation match very well with the observed gaps at 13.2 and 32.3~au.

In Figure \ref{fig:hltau_param} (b), we present the perturbed density distributions with 20, 30, and 40 Earth-mass planets to show how a change in planetary mass affects the locations of secondary/tertiary gaps.
As seen in the figure, the secondary/tertiary gaps open closer to the planet with a larger mass planet, and open farther from the planet with a lower mass planet.
This is because the stronger the spiral arms are, they start to shock disk gas closer to the planet.

Another important factor that determines secondary/tertiary gap depths is the background disk temperature profile since it governs the propagation of waves as well as wavelength of the spiral arms.
In Figure \ref{fig:hltau_param} (c), we present the perturbed density distributions with $20\%$ higher ($(h/r)_p=0.081$) and $20\%$ lower temperatures ($(h/r)_p=0.066$) than our fiducial choice ($(h/r)_p=0.074$).
Note that even with a $\pm20\%$ of change in temperature results in a noticeable change in the gap locations -- the secondary gap location changes by about $\pm4$~au, which is about the linear resolution for the Band 6 ALMA continuum data, and the tertiary gap location changes by about $\pm2$~au.
We also test with different temperature power-law slopes, $q=0.8$ and $q=0.5$, fixing the temperature at the planet's location.
As shown in Figure \ref{fig:hltau_param} (d), we find that the temperature slope can also affect the gap locations although the effect may not be as significant as changing the background temperature.

Because of this degeneracy in gap locations, there are other possible sets of planetary masses and disk temperature profiles that can produce similar ring/gap positions as seen in the HL Tau disk.
For example, when we increase the disk temperature by $20~\%$ we find that the two gap locations can be well reproduced by a 60 Earth-mass planet (see the black dashed curve in Figure \ref{fig:hltau_param} (a)).
Having more accurate observational measurements of the disk temperature structure and other disk properties, such as the disk surface density and viscosity which can affect spiral wave excitation and propagation, will therefore help constrain the mass of the currently unseen planet.
Also, numerical simulations that can simultaneously compute gas and dust dynamics are warranted as the present calculation simulates only the gas component.

With only one planet we were not able to reproduce finer structures (e.g., B3/D3, B4/D4, B7/D7 in \citealt{alma15}).
It is interesting to speculate that two additional newly formed planets in the rings B3 ($\sim47$~au) and B7 ($\sim97$~au) have just started to open the shallow gaps, possibly triggered by the enhancement of solid population at the local pressure maxima created by a hypothesized planet at B5 (and the resulting secondary gap).

\section{DISCUSSION}
\label{sec:discussion}

{\it Observability of the secondary/tertiary gaps.}
In general, $\sim10~\%$ of local density enhancement --  corresponding approximately to $\Delta\Sigma_{\rm gap} \sim 1.2$ -- is known to be able to induce trapping of solid particles in the density rings \citep[e.g.,][]{pinilla12,zhu14,dong17}, although the exact number should be dependent upon the unperturbed density and temperature profiles.
As shown in Figure \ref{fig:long_term}, we find that all secondary/tertiary gaps but the $0.1~M_{\rm th}$ case have $\Delta\Sigma_{\rm gap} \geq 1.2$ after 1000 planetary orbits.
This suggests that even for the planets with $M_p < 1 M_{\rm th}$, sub-Jovian mass in typical protoplanetary disks, the secondary and tertiary rings may collect enough solid particles to produce observable signatures in (sub-)mm continuum, when sufficient time has passed.
It is also possible that the secondary and tertiary gaps are detectable in gas, or scattered light which traces small $\mu$m-sized particles that are well mixed with gas, for high-mass planets.
For instance, \citet{yen16} have detected two gaps in HCO+ emission from the HL Tau disk with gap depths, measured in a similar way to ours, of $\sim 2.4$ and $\sim 5.0$.
As our $3~M_{\rm th}$ case in Figure \ref{fig:long_term} (c) suggests, the secondary and tertiary gap depths for massive planets can be well above the detected gap depths in HL Tau.

{\it Planet formation.}
Solid particles (with appropriate sizes) can be efficiently collected in density rings via aerodynamic drag \citep[e.g.,][]{pinilla12}.
The concentration of particles in the multiple rings formed by planets will provide a favorable condition to expedite the growth of solid bodies within the bumps through coagulation between solid particles, gravitational instability \citep{youdin02}, the streaming instability \citep{youdin05}, and more efficient pebble accretion onto planetesimals \citep{johansen10,ormel10,lambrechts12,morbidelli12}. 
It can also prevent solid particles from rapidly drifting inward, which would happen in a disk with a smooth radial density distribution, thus  potentially overcoming the radial drift barrier \citep{whipple72,weidenschilling77}.
We also note that when the first planet forms in a disk, it can have an influence on the position of subsequent planet formation in the disk. 

{\it Three dimensional effects.}
In three dimensions, it is possible that planet-driven spiral waves can become unstable to the spiral wave instability \citep[SWI;][]{bae16a,bae16b}.
On the other hand, the three-dimensional simulations in \citet{bae16b} indicate that the formation of secondary/tertiary rings and gaps are not completely suppressed even when the SWI is in action -- in their simulation the three gaps open in the inner disk at $R \sim 0.85, 0.6$, and 0.45 (see their Figure 5) and these correspond to the radii at which each of the three arms start to become apparent in the density plot (see their Figure 6). 
It is likely though that the rings and gaps become less prominent than when the SWI is absent. 

{\it Outer disk.}
With a planetary mass range and the disk temperature profiles considered in this work, planets excite only one arm in the outer disk.
On the other hand, more than one spiral arm can excite in the outer disk depending on the planetary mass and the disk temperature profile (Bae \& Zhu 2017, in preparation), in which case more than one gap can open in the outer disk.
Also, as noted in \citet{zhu14}, it is possible that the vortex at the primary ring in the outer disk can excite density waves.
The waves then can steepen to shocks, leading to the formation of additional sets of rings/gaps at larger radii.

%figure 6
\begin{figure}
\centering
\epsscale{1.1}
\plotone{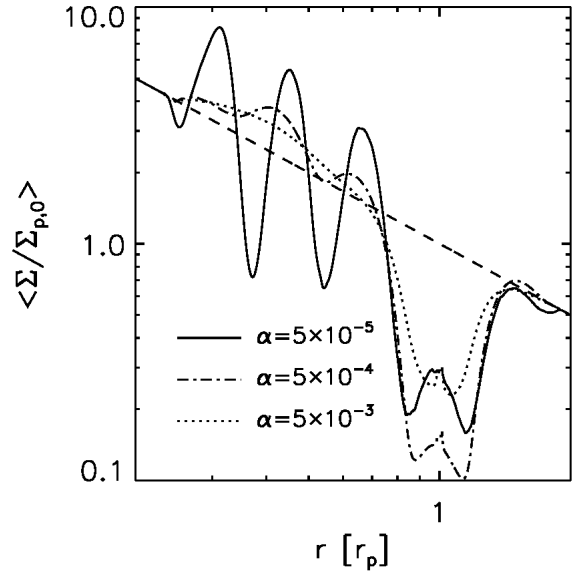}
\caption{Same as Figure \ref{fig:long_term} (a) but with (solid) $\alpha=5\times10^{-5}$, (dashed-dotted) $\alpha=5\times10^{-4}$, and (dotted) $\alpha=5\times10^{-3}$. Note that the secondary and tertiary gaps form with $\alpha=5\times10^{-4}$, with reduced depths compared with the fiducial model, but do not form with $\alpha=5\times10^{-3}$. The density in the corotation region is higher with $\alpha=5\times10^{-5}$ than with $\alpha=5\times10^{-4}$, which is counter-intuitive. This is because the RWI triggers at the outer gap edge with $\alpha=5\times10^{-5}$, but not with $\alpha=5\times10^{-4}$, redistributing gas and (partially) filling the gap.}
\label{fig:viscosity}
\end{figure}

{\it Effect of viscosity.}
The fiducial models in Section \ref{sec:results} used $\alpha=5\times10^{-5}$.
To test the effect of larger viscosity, we adopt $\alpha=5\times10^{-4}$ and $5\times10^{-3}$ for the fiducial model presented in Section \ref{sec:3mth} ($3~M_{\rm th}$ planet case).
As shown in Figure \ref{fig:viscosity}, we find that the secondary and tertiary gaps still open with $\alpha=5\times 10^{-4}$ but their formation is suppressed with $\alpha=5\times10^{-3}$.
If the multiple concentric gaps observed in young protoplanetary disks are due to spiral shocks driven by planets, the existence of those gaps might thus imply low viscosities in the disks ($\alpha \lesssim 10^{-3}$), in potential support of recent theoretical \citep[e.g.,][]{bai13,gressel15,hartmann17} and observational studies \citep[e.g.,][]{flaherty15,flaherty17,pinte16,mulders17}.
We also note that the tertiary gap forms at a slightly smaller radius with $\alpha=5\times 10^{-4}$ than with $\alpha=5\times 10^{-5}$, suggesting that having accurate observational measurments on the disk viscosity is also desired to determine the mass of unseen planets.
A more complete exploration of the parameter space is out of the scope of the paper and will be presented in a future paper; quantitatively, however, additional gaps can open when the ``effective'' viscosity \citep[e.g.,][]{goodman01} arising from spiral shocks is larger than the background disk viscosity.

{\it Orbital eccentricity and migration.}
When a planet has a non-zero orbital eccentricity, it will give a rise to new family of spiral waves having differing pattern speeds around the mean motion of the planet \citep{goldreich80}.
As these spiral waves will become shocks at different radial locations, additional gaps will become broader and shallower. 
Orbital migration can also affect the properties of the gaps (e.g., location, depth) when migration is fast \citep[e.g.,][]{malik15}.
In particular, the gap depths at any instance can be shallower as the gap-opening occurs over an extended period of time (see Figure \ref{fig:long_term}).

{\it Comments on individual disks with multiple gaps.}
We have shown HL Tau as a possible example of a single planet opening multiple sets of rings/gaps.
To help facilitate future numerical modeling of the observed disks with multiple gaps, here we comment on some individual disks.
For the TW Hya disk, it may well be that a low-mass planet ($< 1M_{\rm th}$) embedded in the ring at 40~au opens the double gap at 37 and 43~au via the primary spiral shock, as suggested by \citet{dong17}, and opens a secondary gap at 22~au through a shock from the secondary arm.
This resembles our $0.1~M_{\rm th}$ model.
For the HD~163296 disk, presumably a more massive planet ($\gtrsim 1M_{\rm th}$) opens a broad gap around 160~au and the secondary and tertiary spiral arms open two gaps in the inner disk at 60 and 100~au, similar to what is seen in our $3~M_{\rm th}$ model.

{\it Future work.}
In a forthcoming paper (Bae \& Zhu 2017, in preparation), we discuss the origin of multiple spiral arms driven by a single planet.
Better understanding of the formation of secondary/tertiary arms will enable comparison between linear theory and numerical results.
It will also help make predictions for secondary/tertiary gap locations that can be used as a prescribed formula, for example, in the pebble accretion calculations where the evolution of the gas surface density often relies on simplified viscous evolution (plus a prescribed primary gap formula in some models).
Applications to individual disks with multiple rings/gaps can shed light on the existence of planets in the disks, but also provide critical insights into studies of planet formation and planet-disk interactions.

\acknowledgments

JB is grateful to Ke Zhang for providing the radial intensity profile of the ALMA HL Tau data and for helpful discussion. 
This work was supported in part by NASA grant NNX17AE31G.
We acknowledge the following:  computational resources and services provided by Advanced Research Computing at the University of Michigan, Ann Arbor; the XStream computational resource, supported by the National Science Foundation Major Research Instrumentation program (ACI-1429830); the Extreme Science and Engineering Discovery Environment (XSEDE), which is supported by National Science Foundation grant number ACI-1548562; and the NASA High-End Computing (HEC) Program through the NASA Advanced Supercomputing (NAS) Division at Ames Research Center. 
We made use of the following ALMA data: ADS/JAO.ALMA $\#$2011.0.00015.SV. ALMA is a partnership of ESO (representing its member states), NSF (USA) and NINS (Japan), together with NRC (Canada), NSC and ASIAA (Taiwan), and KASI (Republic of Korea), in cooperation with the Republic of Chile. The Joint ALMA Observatory is operated by ESO, AUI/NRAO and NAOJ. The National Radio Astronomy Observatory is a facility of the National Science Foundation operated under cooperative agreement by Associated Universities, Inc.

\end{document}